\documentclass[12pt,thmsa]{article}
\usepackage{amsfonts}

\input{tcilatex}
\begin{document}

\author{Harald Grosse$^a$ \and Karl-Georg Schlesinger$^b$ \qquad \\
$^a$Institute for Theoretical Physics\\
University of Vienna\\
Boltzmanngasse 5\\
A-1090 Vienna, Austria\\
e-mail: grosse@doppler.thp.univie.ac.at\\
$^b$Erwin Schr\"{o}dinger Institute for Mathematical Physics\\
Boltzmanngasse 9\\
A-1090 Vienna, Austria\\
e-mail: kgschles@esi.ac.at}

\title{A suggestion for an integrability notion for two dimensional spin systems}
\date{UWThPh-2001-12}
\maketitle

\begin{abstract}
We suggest that trialgebraic symmetries might be a sensible starting point
for a notion of integrability for two dimensional spin systems. For a simple
trialgebraic symmetry we give an explicit condition in terms of matrices
which a Hamiltonian realizing such a symmetry has to satisfy and give an
example of such a Hamiltonian which realizes a trialgebra recently given by
the authors in another paper. Besides this, we also show that the same
trialgebra can be realized on a kind of Fock space of $q$-oscillators, i.e.
the suggested integrability concept gets via this symmetry a close
connection to a kind of noncommutative quantum field theory, paralleling the
relation between the integrability of spin chains and two dimensional
conformal field theory
\end{abstract}

\section{Introduction}

We should stress immediately at the beginning that what we discuss on the
following pages is a proposal for an integrability notion for two
dimensional rectangles of spins which we believe contains necessary features
such a notion should have and has, in addition, the merit of a certain
naturality. We do not claim at the present stage that this is the final word
in the important question of integrabilty notions for higher dimensional
spin systems than spin chains. It is an open question if there exists any
sensible intergrability notion in the classical sense of conserved
quantities for such systems. What we show is that there exists a concept -
which one might maybe call ``weak integrability'' - which is linked to a
symmetry structure which relates it to four dimensional topological field
theory and $q$-deformed quantum field theory, much the same way the quantum
group symmetries of integrable spin chain models relate these to three
dimensional topological field theory and two dimensional conformal field
theory.

\bigskip

\section{Spin rectangles}

Trialgebras are algebraic structures with two associative products and a
coassociative coproduct, all linked in a compatible way. Trialgebras relate
to bialgebras as do Hopf algebras to groups. The explicit examples of
trialgebras constructed in \cite{GS} and \cite{GS2} can be seen as a kind of
second quantization of quantum groups. Concretely:

\bigskip

\begin{definition}
A trialgebra $(A,*,\Delta ,\cdot )$ with $*$ and $\cdot $ associative
products on $A$ (where $*$ may be partially defined, only) and $\Delta $ a
coassociative coproduct on $A$ is given if both $(A,*,\Delta )$ and $%
(A,\cdot ,\Delta )$ are bialgebras and the following compatibility condition
between the products is satisfied for arbitrary elements $a,b,c,d\in A$: 
\[
(a*b)\cdot (c*d)=(a\cdot c)*(b\cdot d)
\]
whenever both sides are defined.
\end{definition}

\bigskip

An example where $*$ is partially defined is given by starting from a
bialgebra and passing to its tensor algebra. i.e. taking the tensor product
as the second product. In physics parlance the partial definition of $*$
means that we can only multiply ``$n$ particle'' with ``$m$ particle''
states for $n=m$ but not for $n\neq m$. We will only encounter trialgebras
which are partially defined in precisely this way, in the sequel. In \cite
{GS} we gave a formulation of the above compatibility condition for $\cdot $
and $*$ in terms of $R$-matrices $R_p$ and $R_q$ if both bialgebras, $%
(A,*,\Delta )$ and $(A,\cdot ,\Delta )$, are quantum groups with $R$%
-matrices $R_p$ and $R_q$, respectively. In analogy to quantum groups we
will call trialgebras of this kind \textit{second quantized quantum groups}.

It is generally be viewed an essential feature of integrability for a one
dimensional spin chain if the algebraic structure (Hamiltonian and
symmetries) can systematically be derived for a chain with $n+1$ spins from
one with $n$ spins which is exactly satisfied in the presence of a quantum
groups symmetry (as e.g. in the XXZ model). In this case it is the coproduct
which allows to the data for the longer chain.

Integrability for a two dimensional spin system should then naturally mean
that we can systematically derive the data of larger and larger rectangles
of spins from smaller ones. But this means, especially, that the dimensional
reduction on the two possible directions has to lead to integrable spin
chain models. So, we have two quantum groups for the two integrable spin
chains. In addition, we have to have the above compatibility of the two
products because when enlarging both edges of a spin rectangle the result
should not depend upon the choice which edge we start to enlarge first, i.e.
the two products have to interchange. Since if we enlarge only one edge, we
enlarge already the whole rectangle, the coproduct necessarily has to be one
and the same for both quantum groups. In conclusion, we should get a second
quantized quantum group whenever we have an integrable spin rectangle.

\bigskip

\begin{remark}
Observe that also the structure of a partially defined $*$ product, of the
kind discussed above, has a natural interpretation in this context: It
simply means that we have a rectangle of spins and not some other two
dimensional array of spins.
\end{remark}

\bigskip

If one accepts a second quantized quantum group as a suggestion for an
integrability criterion for spin rectangles, the next question is how one
could construct a Hamiltonian for a spin rectangle which embodies such a
symmetry. As a simple example, we take the trialgebraic deformation $%
\mbox{\v{U}}_{p,q}^F(sl_2)$ of the quantum group $\mbox{\v{U}}_q(sl_2)$
constructed in \cite{GS} (for the definitions and details, see there). The
requirement that dimensional reduction on the two possible directions has to
lead to integrable spin models, leads to a decisive restriction for the
Hamiltonian, too. The Hamiltonian of the XXZ-model is given by 
\[
H_{XXZ}=\sum_i\left( \sigma _i^{+}\sigma _{i+1}^{-}+\sigma _i^{-}\sigma
_{i+1}^{+}\right) +\lambda _q\sum_i\left( \sigma _i^z\sigma
_{i+1}^z-2\right) +\sum_i\left( \sigma _i^z-\sigma _{i+1}^z\right) 
\]
where $\lambda _q$ is a constant involving the deformation parameter $q$,
the index is running over the $n$ elements of the chain and e.g. $\sigma
_i^{+}$ stands for 
\[
\sigma _i^{+}=\mathbf{1}\otimes ...\otimes \mathbf{1}\otimes \sigma
^{+}\otimes \mathbf{1}\otimes ...\otimes \mathbf{1} 
\]
with $\sigma ^{+}$ occuring in the $i$-th place. If we introduce a second
label $j$ for the second direction in the spin rectangle, the complete
Hamiltonian $H$ should have the form 
\[
H=\sum_{i,j}\left( \sigma _{i,j}^{+}\sigma _{i+1,j}^{-}+\sigma
_{i,j}^{-}\sigma _{i+1,j}^{+}\right) +\lambda _q\sum_{i,j}\left( \sigma
_{i,j}^z\sigma _{i+1,j}^z-2\right) +\sum_{i,j}\left( \sigma _{i,j}^z-\sigma
_{i+1,j}^z\right) +\sum_iH_{p,i} 
\]
with the obvious meaning of e.g. $\sigma _{i,j}^{+}$ and $H_p$ the
Hamiltonian for the spin chain in the second direction, involving the
deformation parameter $p$. Dimensional reduction on the $j$-direction, i.e.
taking an inner trace over $i$ (where by an inner trace we mean that in the $%
i$ direction tensor products are converted into matrix products), leads -
modulo a constant factor given by the number of spins in the $i$-direction
and a constant shift - to 
\begin{eqnarray*}
&&\sum_j\left( \sigma _j^{+}\sigma _j^{-}+\sigma _j^{-}\sigma _j^{+}\right)
+\lambda _q\sum_j\sigma _j^z\sigma _j^z+H_p \\
&=&\sum_j\left( \sigma ^{+}\sigma ^{-}+\sigma ^{-}\sigma ^{+}\right)
_j+\lambda _q\sum_j\left( \sigma ^z\right) _j^2+H_p \\
&=&m\left( 1+\lambda _q\right) \mathbf{1+}H_p
\end{eqnarray*}
which as a Hamiltonian is equivalent to $H_p$. So, we get the correct
dimensional reduction, here. Dimensional reduction on the $i$-direction -
taking an inner trace over $j$ - also reproduces $H_{XXZ}$ correctly,
provided the inner trace over $j$ of $H_p$ is a multiple of $\mathbf{1}$. In
order to realize the trialgebraic symmetry, $H_p$ in addition has to commute
with a representation of the associative algebra structure with respect to
the $\cdot $ product of $\mbox{\v{U}}_{p,q}^F(sl_2)$. In conclusion, any
matrix $H_p$ realizing this commutator constraint together with the above
inner trace condition, would lead to realization of the trialgebraic
symmetry on a spin rectangle system and would therefore lead to a system
which naturally suggests itself as an integrable two dimensional spin model.

\bigskip

\begin{remark}
In the case of integrability for spin chains - where the usual quantum group
symmetries arise - there is an important connection to three dimensional
topological quantum field theories which are determined by just the same
quantum groups. This close connection to a topological theory can be seen as
a clear sign of integrability. It is therefore interesting that our proposal
for an integrability notion for spin rectangles, based on trialgebraic
symmetries, leads to a close connection to the algebraic structure of four
dimensional quantum field theories. Trialgebras lead to bialgebra categories
as their categories of representations and with certain types of trialgebras
(the one used below being an example) lead to so called Hopf algebra
categories. But Hopf algebra categories have been shown to be suitable to
generate four dimensional topological quantum field theories (for the
details of the notions and claims of this remark, see \cite{CrFr}, \cite{CKS}%
, \cite{GS}).
\end{remark}

\bigskip

As a concrete example for $H_p$, start from the algebra with respect to the $%
p$-deformed $\cdot $ product given in \cite{GS} by the table 
\[
\begin{array}{llll}
a\cdot b & = & p^{-1} & b\cdot a \\ 
a\cdot c & = & p & c\cdot a \\ 
a\cdot d & = &  & d\cdot a \\ 
b\cdot c & = & p^2 & c\cdot b \\ 
b\cdot d & = & p & d\cdot b \\ 
c\cdot d & = & p^{-1} & d\cdot c
\end{array}
\]
(for the generators of $U_q(sl_2)$, we get a completely similar table, so,
without loss of generality we can - as concerns the $p$-deformed product -
work with the $a,b,c,d$, only). Observe that one can get a representation of
the $a,b,c,d$ in the following way: Let $\left\{ \left| n\right\rangle ,n\in 
\Bbb{N}\right\} $ dentote a basis and $\Lambda $ the shift operator, i.e. 
\[
\Lambda \left| n\right\rangle =\left| n+1\right\rangle 
\]
Let 
\[
a=d 
\]
with 
\[
a\left| n\right\rangle =p^{n-1}\left| n\right\rangle 
\]
and 
\begin{eqnarray*}
b &=&\Lambda ^{-1}a \\
c &=&a\Lambda
\end{eqnarray*}
This defines a representation for the above table. For $p$ a root of unity
this generates a finite dimensional representation. The $R$ matrix
corresponding to the above table is 
\[
R_p=\left( 
\begin{array}{ll}
1 & 0 \\ 
0 & p
\end{array}
\right) \otimes \left( 
\begin{array}{ll}
1 & 0 \\ 
0 & p^{-1}
\end{array}
\right) 
\]
In analogy to the usual quantum group case, we take 
\[
H_p=\sum_j\left( 
\begin{array}{ll}
1 & 0 \\ 
0 & p
\end{array}
\right) _j\otimes \left( 
\begin{array}{ll}
1 & 0 \\ 
0 & p^{-1}
\end{array}
\right) _{j+1} 
\]
then. Here, we assume $p$ to be a root of unity and the length of the chain
to be adjusted in such a way that we have the above finite dimensional
representation on the complete state space. Assuming periodic boundary
conditions for the $j$ direction, $H_p$ commutes with the $a,b,c,d$. Besides
this, the inner contraction of $H_p$ over $j$ is a multiple of 
\[
\left( 
\begin{array}{ll}
1 & 0 \\ 
0 & p
\end{array}
\right) \left( 
\begin{array}{ll}
1 & 0 \\ 
0 & p^{-1}
\end{array}
\right) =\left( 
\begin{array}{ll}
1 & 0 \\ 
0 & 1
\end{array}
\right) 
\]
and all the requirements on $H_p$ are satisfied, therefore. Hence, we have
found a model of a spin plane which realizes one of the trialgebraic
symmetries constructed in \cite{GS}.

\bigskip

\section{$q$-oscillators}

We are now going to show that one can also realize the trialgebra $%
\mbox{\v{U}}_{p,q}^F(sl_2)$ of \cite{GS} on a deformation of the $q$%
-oscillator algebra. We take the following version of the $q$-oscillator
algebra: The algebra $\mathcal{A}_q$ is the complex associative unital
algebra with generators $a,a^{+},q^N,q^{-N}$ and relations 
\begin{eqnarray*}
q^Nq^{-N} &=&q^{-N}q^N=1 \\
q^Na &=&q^{-1}aq^N \\
q^Na^{+} &=&qa^{+}q^N
\end{eqnarray*}
and 
\begin{eqnarray*}
\left[ a,a^{+}\right] _q &\equiv &aa^{+}-qa^{+}a=q^{-N} \\
\left[ a,a^{+}\right] _{q^{-1}} &\equiv &aa^{+}-q^{-1}a^{+}a=q^N
\end{eqnarray*}
(see e.g. \cite{KS}). Let $T$ be an element of the continuous family of
algebra morphisms which exist from the quantum algebra $\mbox{\v{U}}_q(sl_2)$
into the $q$-oscillator algebra $\mathcal{A}_q$ (see \cite{KS}). More
concretely, this family is given by (for $\alpha \in \Bbb{C}$) 
\begin{eqnarray*}
T_\alpha \left( E\right)  &=&a \\
T_\alpha \left( F\right)  &=&a^{+}\left[ N-2\alpha \right] _q \\
T_\alpha \left( K\right)  &=&q^{N-\alpha }
\end{eqnarray*}
where $\left[ x\right] _q$ denotes the usual $q$-number expression for $x$.

\bigskip

\begin{lemma}
The deformation $\mbox{\v{U}}_{p,q}^F(sl_2)$ of $\mbox{\v{U}}_q(sl_2)$
defines a unique Fock space deformation $\mathcal{A}_{p,q}^F$ of $\mathcal{A}%
_q$ such that $T$ extends also to a morphism of the $\cdot $ product and the
two products of $\mathcal{A}_{p,q}^F$ (the one from $\mathcal{A}_q$ and the
deformation of the symmetric tensor product) are compatible.
\end{lemma}

\proof%
By the defining relations of $\mbox{\v{U}}_{p,q}^F(sl_2)$, we get only a
noncommutative behaviour for the exchange of $a$ and $a^{+}$ and for the
exchange of one of these elements with the unit (which is not a unit for the 
$\cdot $ product but only for the $*$ product, as we remarked in \cite{GS}, 
\cite{GS2}). Precisely, we get 
\begin{eqnarray*}
a\cdot 1 &=&p\ 1\cdot a \\
1\cdot a^{+} &=&p\ a^{+}\cdot 1 \\
a\cdot a^{+} &=&p^2\ a^{+}\cdot a
\end{eqnarray*}
The compatibility of the two products follows by calculation from this.%
\endproof%

\bigskip

Observe that - analogous to the fact that $\mathcal{A}_q$ is not a quantum
group - $\mathcal{A}_{p,q}^F$ is not a trialgebra but an algebraic structure
with two compatible products, only. One can imagine $\mathcal{A}_{p,q}^F$ as
a collection (infinite, if $p$ is not a root of unity) of $q$-oscillators
where the $p$-deformation of the tensor product has introduced a certain
kind of interaction between them. In this sense it is a kind of very simple
toy model version of a noncommutative quantum field theory and shows that
trialgebraic symmetries do, indeed, appear in this context. For the simple
trialgebraic deformations with a scalar deformation function constructed in 
\cite{GS} and \cite{GS2}, we can certainly not expect more than a relation
to such toy models. More physically realistic theories will need more
complicated trialgebras to describe their symmetries.

Let us spent a few more remarks on the ``interaction'' introduced by the
statistics of this deformed tensor product (i.e. the $\cdot $ product): We
have 
\[
a\cdot 1\cdot a^{+}=p^4\ a^{+}\cdot 1\cdot a
\]
and more generally 
\[
a\cdot 1\cdot ...\cdot 1\cdot a^{+}=p^{2\left( l-1\right) }\ a^{+}\cdot
1\cdot ...\cdot 1\cdot a
\]
where the number of units in the middle position is $l$. So, for $0<p<1$ we
have an ``interaction'' which is decreasing with increasing separation of
the $q$-oscillators (if we interpret the different positions in formal words
with respect to the $\cdot $ product as a kind of distance).

If we introduce for an $n$ factor expression the notation
\[
a_{i,n}=1\cdot ...\cdot 1\cdot a\cdot 1\cdot ...\cdot 1
\]
if $a$ appears in the $i$-the position, we get - by compatibility of the two
products - the following behaviour under the $*$ product:
\[
a_{i,n}*a_{j,n}^{+}=a_{j,n}^{+}*a_{i,n}
\]
and, in consequence, the rule
\[
a_{i,n}*a_{j,n}^{+}-q\ a_{j,n}^{+}*a_{i,n}=q^{-N}\ \delta _{ij}
\]
for the $*$ product commutator of different $q$-oscillators.

\bigskip

\textbf{Acknowledgements:}

H.G. was supported under project P11783-PHY of the Fonds zur F\"{o}rderung
der wissenschaftlichen Forschung in \"{O}sterreich. K.G.S. thanks the
Deutsche Forschungsgemeinschaft (DFG) for support by a research grant and
the Erwin Schr\"{o}dinger Institute for Mathematical Physics, Vienna, for
hospitality.

\end{document}